\documentclass[a4paper,10pt]{article}          
\usepackage{graphicx}
\usepackage[centertags]{amsmath}
\usepackage{amssymb}

\newcommand{\proarrow}[0]{\rightarrow}

\newcommand{\proname}[2]{#1 \proarrow #2}

\newcommand{\abs}[1]{\left\lvert #1 \right\rvert}

\newcommand{\ih}[0]{ih}
\newcommand{\il}[0]{il}
\newcommand{\dosh}[0]{2h}
\newcommand{\dosl}[0]{2l}
\newcommand{\unoh}[0]{1h}
\newcommand{\unol}[0]{1l}

\begin{document}
\title{\bf{Non-resonant leptogenesis in seesaw models with an almost conserved  $B-L$} \vspace{1cm}}


\author{{\bf J.~Racker}\footnote{racker@ific.uv.es}\\
\small{Depto.\ de F\'{\i}sica Te\'orica and
IFIC, Universidad de
Valencia-CSIC} \\ 
\small{Edificio de Institutos de Paterna, Apt. 22085, 46071 Valencia,
Spain}}
\date{}

\maketitle

\begin{abstract}
We review the motivations and some results on leptogenesis in seesaw models with an almost conserved lepton number. The paper is based on a talk given at the 5th International Symposium on Symmetries in Subatomic Physics, SSP2012.\\

{\bf Preprint}: IFIC/12-58
\end{abstract}
\section{Introduction}
\label{intro}
Leptogenesis is one of the most attractive mechanisms to explain the
origin of the baryon asymmetry of the Universe (BAU)~\cite{fukugita86}. This is so because it arises naturally in simple extensions of the standard model (SM) which
can also explain why the neutrino masses are so tiny. In this mechanism a lepton asymmetry is produced in the out of equilibrium decay of heavy Majorana neutrinos, which is then partially converted into a baryon asymmetry by non-perturbative sphaleron processes (see~\cite{davidson08} for a complete review). 

In the most economical model (type I seesaw) the heavy neutrinos $N_i$ are SM singlets with Majorana masses $M_i$, which only interact with the lepton doublets $\ell_\alpha\; (\alpha=e, \mu, \tau)$ and Higgs field $h$ via Yukawa interactions, $\mathcal{L}_Y = - \lambda_{\alpha i}\,{\widetilde h}^\dag\, \overline{P_R N_i} \ell_\alpha + h.c.\,$.  The number of new parameters associated to the type I seesaw with three singlets, one per each SM family, is 18. However the baryon asymmetry $Y_B \equiv n_B/s$ produced via $N_1$-leptogenesis depends on a few combinations of them (here $n_B$ and $s$ are the baryon and entropy densities). When flavour effects~\cite{barbieri99,endoh03,abada06,nardi06,abada06II,blanchet06} are not relevant the main parameters are $M_1$, which determines the epoch of leptogenesis, $\epsilon_1$, that gives a measure of the amount of CP violation per $N_1$-decay (see below for a precise definition), and the effective mass, $\tilde m_1 \equiv (\lambda^\dag \lambda)_{11} v^2/M_1$ (with $v$ the vev of the Higgs field), which is an appropriate measure of the intensity of the Yukawa interactions of $N_1$. If the CP asymmetry $\epsilon_1$ is constant during leptogenesis (which is usually a good approximation), the final baryon asymmetry, $Y_B^f$, is simply proportional to $\epsilon_1$. In this case it can be expressed as $Y_B^f = k \epsilon_1 \eta$, with $k \simeq 1/724$ a numerical factor and $\eta$ is the so called efficiency, which carries the dynamical information and it is mainly a function of $\tilde m_1$. By definition $\abs{\eta} \le 1$ and the maximum efficiency is obtained when $\tilde m_1 \sim 10^{-3}$~eV. This value is determined by the condition that the decay rate of $N_1$ equals the Hubble expansion rate at a temperature $T=M_1$, so that the Yukawa interactions of $N_1$ are barely out of equilibrium at the time it becomes non-relativistic. This result is amazing given that the contribution of $N_1$ to the masses of the light neutrinos ($m_i, i=1,2,3$) in the type I seesaw is expected to be of the same order as $\tilde m_1$ (barring cancellations due to phases). In other words, an efficient leptogenesis mechanism suggests a scale for the light neutrino masses which is roughly of the correct order of magnitude.  Moreover, the most simple models for leptogenesis require an interesting upper bound for the masses of the light neutrinos, $m_i \lesssim 0.15$~eV in the one flavour approximation~\cite{buchmuller03} and $m_i \lesssim$~few eV when flavour effects are taken into account~\cite{abada06,desimone07II}. 

The conditions described above provide a highly non-trivial connection between baryogenesis via leptogenesis and the low energy parameters $m_i$. But this is not enough at all to probe leptogenesis. Unfortunately in the most simple models no more generic relations can be established between leptogenesis and low energy parameters. In fact, for a very hierarchical spectrum of heavy singlet neutrinos $M_1 \ll M_2 \ll M_3$, 
the $L$-violating CP asymmetry generated in the decay of the lightest singlet 
has an upper bound proportional to $M_1$, the so-called  
Davidson-Ibarra (DI) bound \cite{davidson02}. 
This implies a lower bound $\sim 10^9$~GeV for the mass of the sterile neutrinos
in order for $N_1$-dominated leptogenesis to be successful.
Careful numerical studies show that the DI bound can be evaded for moderate 
hierarchies, e.g. the lower bound on $M_1$ is relaxed by more than one order
of magnitude with respect to the hierarchical limit one for $M_3/M_2 \sim
M_2/M_1 \sim 10$~\cite{hambye03}. 
However  to reach these low values of $M_1$ some unlikely 
cancellations are needed, which are not motivated by 
any underlying symmetry. 
 Flavour effects do not substantially change 
 this result. Therefore leptogenesis occurs at very high energies in these scenarios. In addition no generic relation can be made between low and high energy phases, i.e. leptogenesis can work for any value of the observable PMNS phases~\cite{davidson07}.

In conclusion, leptogenesis in the context of the type I seesaw with hierarchical heavy neutrinos provides a simple and natural explanation to the BAU, but it will not be possible to test this scenario in foreseeable experiments. This has motivated research in different directions. For example in~\cite{frere08} some ways to falsify (rather than probe) leptogenesis at the LHC were investigated.  Also, one can avoid the DI bound resorting to resonant leptogenesis, i.e., a resonant 
enhancement of the CP asymmetry which occurs when there are at least  two strongly 
degenerated heavy neutrinos, such that $M_2 - M_1 \sim \Gamma_N$, being 
$\Gamma_N$ their decay width~\cite{covi96II,Anisimov05}. In this scenario, 
leptogenesis is feasible at much lower temperatures, $T \sim {\cal O}(1 \;\rm{TeV})$
\cite{pilaftsis04,piu,pilaftsis08,deppisch10}. However it is not enough to have leptogenesis at the TeV scale in order to probe it. This is so because the most crucial parameters for observing effects from the heavy neutrinos are the active-sterile neutrino mixings, which in the type I seesaw are roughly given by the ratio $m_D/M_1 \sim \sqrt{m_i/M_1}$ (with $m_D \sim \lambda_{\alpha i} v$), and hence are too small. 

Therefore it is very interesting that there are well motivated seesaw models which not only yield a heavy neutrino quasi-degenerate spectrum but can also provide a large active-sterile neutrino mixing, namely those 
that have an approximately conserved $B-L$~\cite{mohapatra86} (with $L$ being conserved at the perturbative level).
In these models 
 the tiny neutrino masses are proportional to  small 
lepton number-breaking parameters, which are technically natural 
since a larger symmetry is realized when they vanish. This implies that the heavy neutrinos can be much lighter than in the generic seesaw, within the energy 
reach of LHC.
Also, lepton flavour violating rare decays as well as non-unitarity of the leptonic mixing matrix
are present even in the limit of conserved $B-L$, and therefore they are unsuppressed
by the light neutrino masses \cite{Bernabeu87,GonzalezGarcia91,Hernandez09}.
As a consequence, 
much attention has been devoted recently to this class of low scale seesaw models,  
since they have a rich phenomenology both at  LHC  \cite{Han06,delaguila07,Kersten07}
and at  low energy charged lepton rare decay experiments, such as 
$\mu \rightarrow e \gamma$, and also lead to successful resonant leptogenesis 
\cite{asaka08}.

It has also been noticed that 
even if the heavy neutrinos that generate the BAU are not quasi Dirac, 
or the mass splitting is outside the resonant regime, in seesaw models with almost conserved 
$B-L$ the scale of leptogenesis can be lower than in the standard seesaw  
\cite{antusch09,racker12}, provided flavour effects are at work. 
This is so because there is a $L$-conserving part in the flavoured  CP-asymmetries 
 which escapes the DI bound. In these notes we review and summarize the results of~\cite{racker12} on the possibility of having successful leptogenesis driven by the 
purely flavoured  $L$-conserving contribution to the CP asymmetries, in the context 
of seesaw models with small violation of $B-L$.

\section{Leptogenesis in models with an almost conserved $B-L$}
\label{sec:dos}

In general there are at least two species of neutrinos involved in leptogenesis, one, called here $N_1$, which is mainly responsible for the generation of the lepton asymmetry during its production and decay, and another one, $N_2$, that makes the most important virtual contribution to the CP asymmetry in $N_1$ decays. If $B-L$ is only slightly violated, then each $N_i$ must satisfy one of the two following conditions:
\begin{itemize}
\item[(i)] $N_i$ is a Majorana neutrino with two degrees of freedom, whose Yukawa interactions violate lepton number and therefore the couplings $\lambda_{\alpha i }$ must be small.
\item[(ii)] The $N_i$ is a Dirac or quasi-Dirac neutrino with four degrees of freedom;
this means that
  there are two Majorana neutrinos $N_{\ih}$ and $N_{\il}$ with masses $M_i + \mu_i$ and $M_i - \mu_i$ respectively.  The parameter $\mu_i \ll M_i$ measures the amount of $B-L$ violation,
  so that if $B-L$ is conserved, $\mu_i =0$ and $N_i= (N_{\ih} + i N_{\il})/\sqrt{2}$ 
  is a Dirac fermion. 
  The Yukawa interactions can be expressed as
\begin{equation}
\mathcal{L}_{Y_{Ni}} = - \lambda_{\alpha i}\,{\widetilde h}^\dag\,
\overline{P_R \frac{N_{\ih} + i N_{\il}}{\sqrt{2}}} \ell_\alpha  -
  \lambda^\prime_{\alpha i}\,{\widetilde h}^\dag\,
\overline{P_R \frac{N_{\ih} - i N_{\il}}{\sqrt{2}}} 
  \ell_\alpha + h.c. \, ,
\end{equation}
where 
$\lambda^\prime_{\alpha i} \ll 1$. 
The terms proportional to $\lambda^\prime_{\alpha i}$ induce lepton number violation even when $\mu_i \to 0$ and hence they are similar in nature to the ones described in (i). Instead the $\lambda_{\alpha i}$ can be large, because they do not vanish in the $B-L$ 
conserved limit: in the absence of $\mu_i$ and 
$\lambda^\prime_{\alpha i}$, a perturbatively conserved lepton number can be defined, 
by assigning $L_N =1$  to $N_i$, and $L_{\ell_\alpha} =1$ to the SM leptons.

 There are two cases that are relatively easy to analyze,
\begin{itemize}
\item (iia) $\mu_i \ll \Gamma_{N_{\ih}}, \Gamma_{N_{\il}} \; \,$  (Dirac limit), and 
\item (iib) $\Gamma_{N_{\ih}}, \Gamma_{N_{\il}} \; \ll \mu_i \ll M_i \; \,$  (Majorana limit). 
\end{itemize} 
Here $\Gamma_{N_{\ih}}$ and $\Gamma_{N_{\il}}$ are the decay widths of $N_{\ih}$ and $N_{\il}$ respectively.
\end{itemize}

A comprehensive research on leptogenesis in models with small violation of $B-L$ can be obtained considering the different possibilities (i) or (ii) for both $N_1$ and $N_2$. Since we have not considered the widely-studied case of a resonant contribution of $N_2$ to the CP asymmetry in $N_1$ decays, the optimum situation for generating a lepton asymmetry is when $N_2$ satisfies (ii). In this way the CP asymmetries $\epsilon_{\alpha 1}$ in the decays of $N_1$ into leptons of flavour $\alpha$, $\epsilon_{\alpha 1} \equiv \tfrac{\Gamma(\proname{N_1}{\ell_\alpha h}) - \Gamma(\proname{N_1}{\bar \ell_\alpha \bar h})}{\sum_\alpha \Gamma(\proname{N_1}{\ell_\alpha h}) + \Gamma(\proname{N_1}{\bar \ell_\alpha \bar h})}$, being proportional to the Yukawa couplings of $N_2$, can be enhanced. In turn, for $N_1$ the simplest possibility is (i). It can  also satisfy (iib), in which case $N_{\unol}$ and $N_{\unoh}$ behave as two independent Majorana neutrinos regarding the generation of the BAU, that would roughly double with respect to case (i). However if  $N_1$ satisfies (iia), then it is (or effectively behaves as) a Dirac neutrino, i.e. lepton number is conserved in its decays, and therefore the only possibilities to end up with a non-zero BAU is to have important washouts from the two Majorana components of 
$N_2$ (if $\mu_2 \gg \Gamma_{N_{\dosl}, N_{\dosh}}$) or let the sphalerons freeze out during leptogenesis~\cite{gonzalezgarcia09}.

Motivated by the previous discussion we have considered a scenario for leptogenesis involving three fermion singlets $N_1, N_{\dosl}, N_{\dosh}$ (each of them having two degrees of freedom), with respective masses $M_1, M_2 - \mu_2, M_2 + \mu_2$ and Yukawa couplings given by the Lagrangian
\begin{equation}
\label{eq:lagy12}
\mathcal{L}_{Y} = - \lambda_{\alpha 1}\,{\widetilde h}^\dag\, \overline{P_R N_{1}} \ell_\alpha - \lambda_{\alpha 2}\,{\widetilde h}^\dag\,
\overline{P_R \frac{N_{\dosh} + i N_{\dosl}}{\sqrt{2}}} \ell_\alpha  + h.c. \; .
\end{equation}
The parameters $\lambda_{\alpha 1}$ violate lepton number and hence $\lambda_{\alpha 1} \ll \lambda_{\alpha 2}$. 

As shown in~\cite{racker12} it is convenient to take $M_1 < M_2$ in order to obtain the lowest energy scale for leptogenesis within this framework, which corresponds to the so called $N_1$-leptogenesis. Then $Y_B^f$ is  proportional to the CP asymmetries $\epsilon_{\alpha 1}$, which have a $L$-violating part suppressed by the small $L$-violating parameter $\mu_2$ and an unsuppressed $L$-conserving piece, $\epsilon_{\alpha 1}^L$, whose contribution to the total CP asymmetry is null, i.e. $\sum_\alpha \epsilon_{\alpha 1}^L = 0$. In order to have a large $Y_B^f$ (not suppressed by $\mu_2$), it is mandatory to have flavour effects so that there is a contribution to $Y_B^f$ coming from $\epsilon_{\alpha 1}^L$~\cite{nardi06}. In turn, to have the appropriate flavour effects, it is crucial to demand that the couplings of $N_1$ and $N_2$ be small enough, such that the Yukawa interactions of the $\tau$ are the strongest ones (see~\cite{racker12} for a detailed explanation). When this happens the density matrix of leptons is diagonal in the orthogonal  basis $(\ell_\tau, \ell_{\tau \perp},{\ell'}_{\! \tau \perp})$, with $\ell_{\tau \perp}$ and ${\ell'}_{\! \tau \perp}$ being determined by the fastest interaction acting in the plane perpendicular to $\ell_\tau$. Something similar occurs in the antilepton sector. Then, as a first approximation, the lepton asymmetries in the flavours $\ell_\tau, \ell_{\tau \perp}$, and ${\ell'}_{\! \tau \perp}$ evolve independently. In this case, although $\sum_\alpha \epsilon_{\alpha 1}^L = 0$, $Y_B^f$ can get contributions from the individual $\epsilon_{\alpha 1}^L$. This is so because the final amount of lepton asymmetry in a given flavour also depends on how much of the produced asymmetry was erased, and this can be different for each flavour.

Actually the evolutions of the different lepton flavour asymmetries are not completely independent. On one hand, spectator processes~\cite{buchmuller01,nardi05} effectively couple the flavour asymmetries, nevertheless we have checked that their effect on $Y_B^f$ is at most a few tens of percent. One the other hand, there are  $L$-conserving but
$L_\alpha$-violating scatterings $\proname{\ell_\beta h}{\ell_\alpha h}$, 
$\proname{\ell_\beta \bar{h}}{\ell_\alpha \bar{h}}$, and
$\proname{h\bar{h}}{\ell_\alpha\bar{\ell_\beta}}$, hereafter called generically
flavour changing interactions (FCI), which are inherent to models with an approximately conserved $B-L$.
The FCI play a crucial role because they tend to equilibrate the different flavour asymmetries~\cite{aristizabal09,fong10}, effectively diminishing flavour effects and consequently $Y_B^f$. The cross sections of the FCI have been calculated in~\cite{racker12}, finding important differences with previous literature.

Summarizing, in order to determine the BAU generated in models for leptogenesis with small violation of $B-L$, it is very important to consider the FCI and the adequate conditions for having flavour effects. The set of Boltzmann equations (BE) taking into account these elements can be read in~\cite{racker12}. For the case $\mu_2 \gg \Gamma_{N_{\dosl, \dosh}}$ the BE are like the ones typically found in the literature. Instead, if $\mu_2 \ll \Gamma_{N_{\dosl, \dosh}}$ then $N_{\dosl}$ and $N_{\dosh}$ combine to form a Dirac neutrino $N_2 \equiv (N_{\dosh} + i N_{\dosl})/\sqrt{2}$, and therefore there is an asymmetry generated among the  degrees of freedom of $N_2$ which has to be taken into account~\cite{gonzalezgarcia09}.

\section{Results}
\label{sec:res}
\begin{figure}
  \includegraphics[width=0.35\textheight,angle=270]{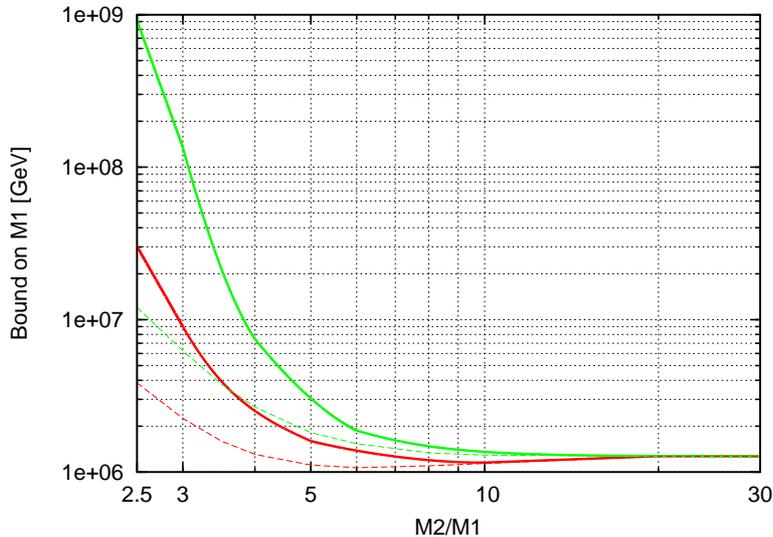}
\caption{Lowest value of $M_1$ yielding successful leptogenesis as a function of $M_2/M_1$. The red curves are for the case $\mu_2 \gg \Gamma_{N_{\dosl, \dosh}}$ and the green ones for $\mu_2 \ll \Gamma_{N_{\dosl, \dosh}}$. The thick continuous curves give the physically correct bound, while the thin dashed ones show the result that would be obtained if the Yukawa couplings of $N_2$ were allowed to take values as large as 1 for all values of $M_2/M_1$.}
\label{fig:1}       
\end{figure}

The relevant parameters for leptogenesis are $M_1$, $M_2/M_1$, $(\lambda^\dag \lambda)_{11}$, $(\lambda^\dag \lambda)_{22}$, the projectors $K_{\alpha i} \equiv \lambda_{\alpha i} \lambda_{\alpha i}^* /(\lambda^{\dagger} \lambda)_{ii}$, and $\mu_2$.
We have determined the minimum value of $M_1$ compatible with successful leptogenesis as a function of $M_2/M_1$, maximizing $Y_B^f$ over the remaining parameters. To obtain the baryon asymmetry we have solved numerically the appropriate set of BE~\footnote{For simplicity we have neglected spectator processes during leptogenesis and the asymmetry developed among the degrees of freedom of the Higgs~\cite{buchmuller01,nardi05}, as well as $\Delta L=1$ scatterings~\cite{nardi07II,fong10II}. However we have checked that their inclusion modifies the results by at most a few tens of percent.}, and to get successful leptogenesis we have required
$Y_B^f  = 8.75 \times 10^{-11}$ \cite{komatsu10}.
The result is represented with the thick continuous curves in Fig.~\ref{fig:1}, the red line corresponding to the case $\mu_2 \gg \Gamma_{N_{\dosl, \dosh}}$ and the green one to $\mu_2 \ll \Gamma_{N_{\dosl, \dosh}}$. 

As can be seen it is possible to have neutrino masses as low as $M_1 \sim 10^{6}$~GeV, i.e. around three orders of magnitude below the lower bound for the standard case of type I seesaw with hierarchical heavy neutrinos. Such lower bound on $M_1$ in turn yields a lower bound for the reheating temperature, $T_{RH}$,  of the same order, since to thermally produce the neutrinos 
$M_1 \lesssim 5 \,  T_{RH}$~\cite{giudice04,buchmuller04,racker08}. An interesting consequence is that the bound $T_{RH} \gtrsim 10^6$~GeV can be compatible with the upper bound on $T_{RH}$ required to avoid the gravitino problem in SUGRA models~\cite{Khlopov84,Ellis84,Kawasaki08}. Moreover, $M_1$ values around $10^{6}$~GeV can be achieved for a wide range of $N_2$ masses and also for different values of the Yukawa couplings (see~\cite{racker12} for details on this point as well as on the relation between the parameters defined above and the light neutrino masses). 

An important issue for obtaining the bound on $M_1$ has been to determine how large the  Yukawa  couplings of $N_2$ can be without violating the condition that the rates of processes involving $N_2$ be slower than the rates of the $\tau$-Yukawa interactions.
For comparison we have also plotted in Fig.~\ref{fig:1} the -wrong- bound that would be obtained if $(\lambda^\dag \lambda)_{22}$ were allowed to be as large as 1. It is clear that as $M_2$ approaches $M_1$ the requirement of an upper bound for $(\lambda^\dag \lambda)_{22}$ becomes very relevant.

Finally let us comment that for simplicity the results depicted in Fig.~\ref{fig:1} were obtained assuming that $\ell_e$ is perpendicular to the decay eigenstates of $N_1$ and $N_2$, so that only two flavour asymmetries are generated. We have checked that in the more general three flavour case it is possible to lower the bound on $M_1$ by a factor up to almost 4 with respect to the two flavour case~\cite{racker12}. 

\section{Conclusions}
Seesaw models with an almost conserved $B-L$ are an interesting alternative to explain the smallness of neutrino masses because they can lead, in principle, to large active-sterile neutrino mixings. We have found another merit of these models, namely that leptogenesis is possible for 
$M_1 \gtrsim 3-10 \times 10^5 \,$~GeV, i.e. around three orders of magnitude below the standard type I seesaw case, without resorting to the resonant enhancement of the CP asymmetry for strongly degenerate heavy neutrinos. 
However, it is also clear that such energy scale is too large to have both, successful non-resonant leptogenesis and active-sterile neutrino mixings large enough to produce observable effects.
\subsection*{Acknowledgments}
The author wishes to thank his collaborators on this subject, Concha Gonz\'alez-Garc\'\i a, Manuel Pe\~na and Nuria Rius. \\
This work has been supported by the Spanish MINECO Subprogramme Juan de la Cierva and it has also been partially supported by the Spanish MINECO grants FPA-2007-60323, FPA2011-29678-C02-01, and 
Consolider-Ingenio CUP (CSD2008-00037). In addition we acknowledge partial support from the  European Union FP7  ITN INVISIBLES (Marie Curie Actions, PITN- GA-2011- 289442). 


%
%


\end{document}